\documentclass[amsmath,amssymb,superscriptaddress]{revtex4-2}

\usepackage[english]{babel}
\usepackage[letterpaper,top=2cm,bottom=2cm,left=3cm,right=3cm,marginparwidth=1.75cm]{geometry}
\emergencystretch=\maxdimen
\usepackage{amsmath}
\usepackage{graphicx}
\usepackage{amsmath, amssymb, graphics, setspace}
\newcommand{\mathsym}[1]{{}}
\newcommand{\unicode}[1]{{}}

\usepackage[colorlinks=true, allcolors=blue]{hyperref}
\usepackage{caption}
\usepackage{subcaption}
\usepackage{natbib}
\usepackage{newtxtext}
\usepackage{newtxmath}

\newcommand\Rey{{\mathrm{Re}}}

\usepackage{color,soul}

\begin{document}
\title{Newtonian fluid dynamics in a misaligned parallel-plate rheometer}
\author{Jian Teng}
\affiliation{Center for Fluid Mechanics, Brown University, Providence, Rhode Island 02912, USA}
\author{Sungwon La}
\affiliation{Center for Fluid Mechanics, Brown University, Providence, Rhode Island 02912, USA}
\author{Jesse T. Ault}
\email{jesse\_ault@brown.edu}
\affiliation{Center for Fluid Mechanics, Brown University, Providence, Rhode Island 02912, USA}

\begin{abstract}
A parallel-plate rotational rheometer measures the viscosity of a fluid by rotating the top plate relative to the bottom plate in order to induce a shear on the fluid and measuring the torques and forces that result as a function of the induced rotation rate. Manufacturing imperfections can often lead to unintentional misalignment of the plates of the rheometer, where the top and bottom plates are not perfectly parallel, and this misalignment can affect the fluid dynamics inside the rheometer. This study examines the effect that misalignment has on the viscosity measurements of Newtonian fluids in the limit of small rheometer gap heights. A theoretical model for the behavior of a general Newtonian fluid in a misaligned rheometer with a small gap height is derived using perturbation expansions. The theoretical results show that at small gap heights, misalignment can produce additional secondary velocity components and pressures in the fluid, which affect the forces and moments in the rheometer. In such cases at small Reynolds numbers, the dominant forces and moments acting on the top plate of the rheometer are the viscous force in the direction parallel to the tilt axis, the pressure moment in the direction perpendicular to the tilt axis and in the cross-sectional plane, and the viscous moment in the direction along the height of the rheometer. These forces and moments on the top plate were found to increase as the misalignment tilt angle increased, leading to an increase in the error of viscosity measurement by the rheometer. Three-dimensional numerical simulations validate the theoretical predictions.
\end{abstract}

\maketitle

\section{Introduction}
A rheometer is a scientific device used to measure the rheological properties of viscous fluids, which are the properties that describe how fluids flow and deform in response to applied forces. The rheometer applies a certain amount of force, stress, or strain to the fluid of interest and measures the response of the fluid in order to understand and derive various fluid properties, including viscosity, elasticity, and viscoelasticity \citep{malkin2022rheology}. Over the last several decades, rheometry has been used to characterize the viscosities and dynamic reactions of various fluids found in everyday use, including those in food, cosmetics, pharmaceuticals, construction materials, adhesives, and household products \citep{singh1999prediction, coussot2005rheometry, gallegos1999rheology, cardoso2015parallel, ahmed2018advances, goh2010tribological, jones1997textural, navaneethan2005application, park2010rheological}. Rheometry also has numerous applications in biology, such as in studying the dynamic behavior of blood under different flow conditions \citep{ihm2020viscosity}, the rheological properties of cells \citep{desprat2006microplates}, and the rheological changes in proteins during alteration \citep{choi2010microfluidic}. 

There are several different types of rheometers, including rotational rheometers, capillary rheometers, vane rheometers, cone and plate rheometers, sliding plate rheometers, and more \citep{covas2000rheological, ferraris2003relating, giacomin1989novel, magnin1990cone}. Each type of rheometer is suited for different fluids and specific applications and has its own unique mechanism by which it creates stress on a fluid and measures its response. The parallel-plate rotational rheometer, which is among the most commonly used types of rheometers, is a device in which a fluid is placed in between two parallel plates, and one of the plates rotates relative to the other in order to induce a shear on the fluid of interest. By measuring the torques and forces applied on the plates by the fluid as a function of the rotation rate of the moving plate, the viscosity of the fluid can be characterized \citep{malkin2022rheology}. The parallel-plate rheometer is widely used in studies due to its flexibility in controlling the gap height in which the fluid is placed, as well as the feature of being able to confine the fluid in a spatially uniform manner \citep{ault2023viscosity}. 

Many studies and experiments are conducted under the assumption that the rheometer is manufactured perfectly without any geometrical imperfections. However, mechanical uncertainties can certainly be introduced through small errors in the manufacturing process. Previous literature has shown that uncertainties in perfectly measuring the actual height of the gap \citep{davies2005gap}, roughness on the surfaces of the plates \citep{carotenuto2013use}, and underfilling or overfilling of the fluid in the rheometer gap \citep{hellstrom2014errors, cardinaels2019quantifying} can lead to significant errors in the resulting measurements of viscosity by the rheometer. Nonparallelism of the plates has also been investigated in several studies due to the significant error it can contribute during experimental measurements, as in the work by \citet{davies2008thin}, the effect of misaligned plates was found to typically lead to a gap height variation error of up to 50 $\mu m$ across the rheometer. This variation is especially important because in the limit of small gap heights, the reported viscosity of fluid samples typically decreases with gap height \citep{kramer1987measurement, connelly1985high, macosko1994rheology, walters1992recent}. When working with a rheometer with a small gap height, misalignment of the plates can introduce significant gap height variations across the rheometer, which can result in an inaccurately reported value for viscosity as evidenced by the gap-dependent nature of the apparent viscosity measurements. 

A recent paper by \citet{ault2023viscosity} further examined the phenomenon of inaccurate viscosity readings due to misaligned plates of a rheometer in the limit of small gap heights. In their study, \citet{ault2023viscosity} performed viscosity measurements of glycerol using a stress-controlled parallel plate rheometer. The hygroscopic nature of the glycerol allows for the absorption of water vapor from the atmosphere, which was found to affect the transient viscosity measurements. However, experimental results from \citet{ault2023viscosity} showed unexpected behavior in rapidly decreasing viscosity at very small rheometer gap heights, which did not match the numerical simulations nor the theoretical behavior derived for this setup based on the assumption of parallel plates. The rate of decrease was dependent on the gap thickness of the rheometer setup, and this relationship was observed to be nonmonotonic. \citet{ault2023viscosity} concluded that this phenomenon observed at small rheometer gap heights for glycerol is due to unintentional misalignment of the rheometer plates. In the regime of extremely small gap heights, slight misalignment can result in additional secondary flows that affect the typical secondary radial recirculation in the rheometer, which would affect the rate of water vapor absorption and thus influence the viscosity readings of the glycerol.

Expanding upon the work done by \citet{ault2023viscosity}, this study focuses on the fluid dynamics inside a misaligned parallel-plate rheometer for general Newtonian fluids and the effect that misalignment has on the resulting viscosity measurements. The effect of plate misalignment has been explored in previous literature; for example, techniques to estimate the misalignment angle have been developed using numerical calculations from torque and force measurements \citep{andablo2010method} and by monitoring ultrasound pulse time-of-flight echoes reflected by the misaligned plate \citep{rodriguez2013using}. Additionally, \citet{clasen2013self} developed a design that uses hydrodynamic lubrication forces to self-align the plates of a rheometer in a parallel manner. Also, a previous study by \citet{andablo2011nonparallelism} derived a numerical solution to the effect of misalignment on the forces applied by Newtonian fluids on a parallel-plate rheometer using a finite-differences numerical method. However, none of the previous studies have developed a full theoretical explanation of the Newtonian fluid dynamics within a misaligned rheometer, nor have they derived a theoretical relationship between the degree of misalignment and the resulting viscosity measurements. In this study, we develop a novel theoretical solution for Newtonian fluid dynamics within a misaligned parallel-plate rheometer. In Section \ref{theory}, we solve for the fluid field within the system from the governing equations and calculate the force and moment acting on the top plate of the rheometer, from which we can predict the resulting viscosity of the fluid. In Section \ref{numerical}, we use full three-dimensional numerical simulations to validate our theoretical predictions and provide additional insight into the system. In Section \ref{result}, we analyze the theoretical and numerical results to identify the dominant forces and moments in the rheometer, and characterize the relationship between plate misalignment and the resulting viscosity measurements of Newtonian fluids.

\section{Theoretical derivation}\label{theory}
\begin{figure}
    \centering
    \includegraphics[width=0.8\textwidth]{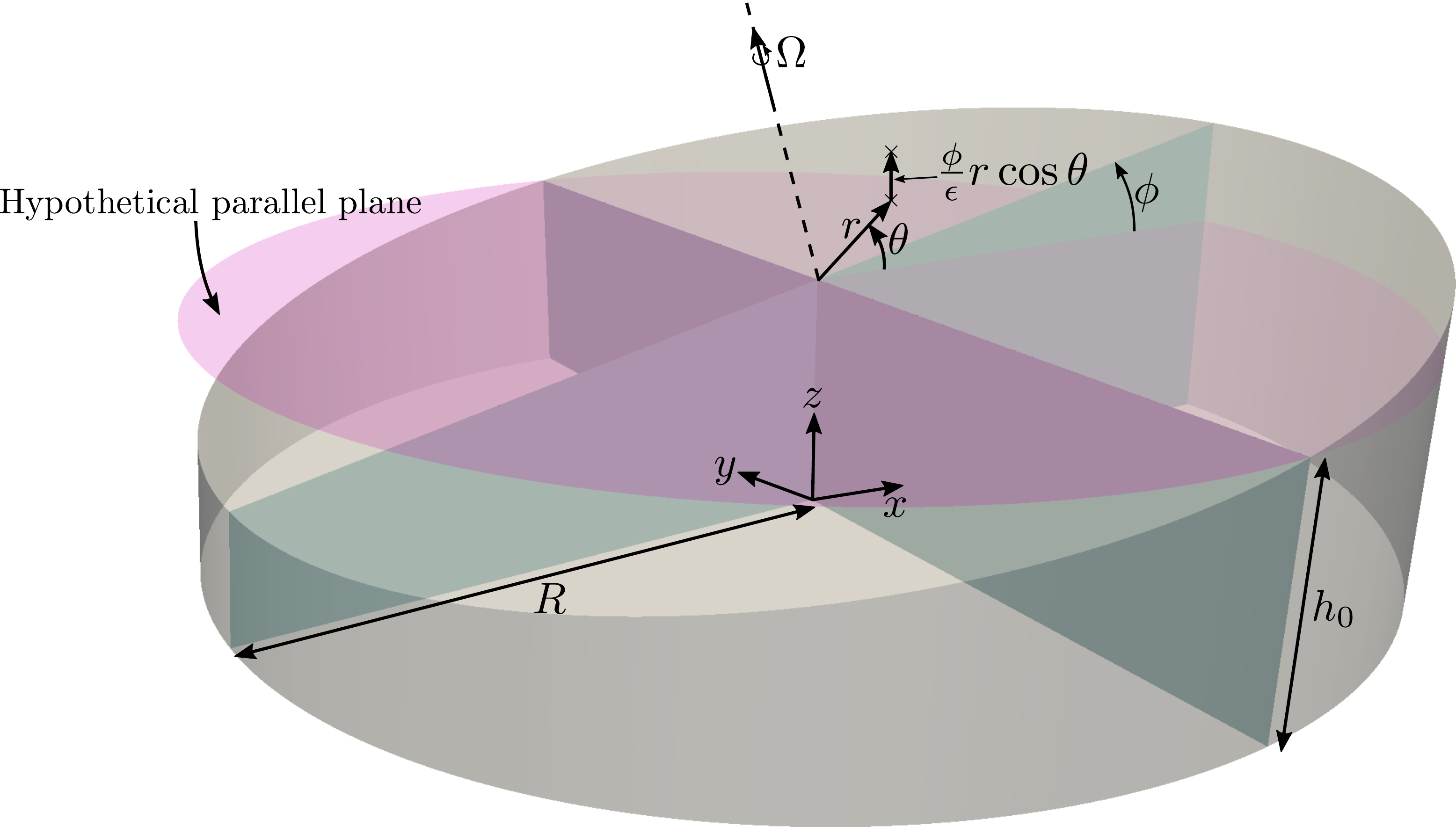}
    \caption{Geometry and coordinate system of the parallel-plate rheometer. The lower plate is stationary, and the upper plate, which is misaligned by a small deflection angle $\phi$, rotates at an angular speed $\Omega$. The radius of the plates and mean gap thickness between the plates are $R$ and $h_0$, respectively. (This figure is reused from \citet{ault2023viscosity})}
    \label{fig:geo}
\end{figure}

In this section, we model the fluid dynamics for a Newtonian fluid in a parallel-plate rheometer with slightly misaligned plates. The parallel-plate rheometer has a radius $R$ and a gap thickness $h(r,\theta,\phi)$ where $(r,\theta)$ are the cylindrical coordinates. The rheometer's upper plate is slightly tilted by a small angle $\phi$, and the upper plate rotates at a rotation speed $\Omega$ while the lower plate remains stationary. The coordinate system and problem setup are shown in Figure \ref{fig:geo}. 
In this paper, we only consider the viscosity measurement of single-phase Newtonian fluids. The governing equations for the fluid in the rheometer are the incompressible Navier-Stokes and continuity equations, which are given by
\begin{equation}\label{eq:variableNS}
\nabla^*\cdot\boldsymbol u^* = 0     ~~~~\text{and}~~~~      \rho\frac{D\boldsymbol{u}^*}{Dt^*}=-\nabla^* p^* + \mu {\nabla^*}^2 \boldsymbol u^*.
\end{equation}
The stars * denote dimensional variable quantities, and $\boldsymbol{u}$, $p$, $\rho$, and $\mu$ are the fluid velocity vector, pressure, density, and viscosity, respectively. The governing equations, which are represented in cylindrical coordinates, can be nondimensionalized with
\begin{equation}\label{eq:scalings}
r=\frac{r^*}{R},~~~z=\frac{z^*}{h_0},~~~u_r=\frac{u_r^*}{\Omega R},~~~u_\theta=\frac{u_\theta^*}{\Omega R},~~~u_z=\frac{u_z^*}{\Omega h_0},~~~p=\frac{p^*}{\mu \Omega R^2/h_0^2},
\end{equation}
where $h_0$ is the nominal gap height for $\phi=0$. With these nondimensional parameters, the nondimensional Navier-Stokes equations in cylindrical coordinates become
\begin{subequations}\label{eq:variableNS_cylindrical}
\allowdisplaybreaks
\begin{align}
&\frac{1}{r}\frac{\partial}{\partial r}\left(r u_r\right) + \frac{1}{r}\frac{\partial u_\theta}{\partial \theta} + \frac{\partial u_z}{\partial z} = 0,\\
%
%
&\Rey\left(u_r\frac{\partial u_r}{\partial r}+\frac{u_\theta}{r} \frac{\partial u_r}{\partial \theta}-\frac{u_\theta^2}{r}+u_z\frac{\partial u_r}{\partial z}\right)=-\frac{\partial p}{\partial r}+\frac{\epsilon^2}{r}\frac{\partial}{\partial r}\left(r\frac{\partial u_r}{\partial r}\right)-\epsilon^2\frac{u_r}{r^2}+\frac{\epsilon^2}{r^2}\frac{\partial^2 u_r}{\partial \theta^2}-\frac{2\epsilon^2}{r^2}\frac{\partial u_\theta}{\partial \theta}+\frac{\partial^2 u_r}{\partial z^2},\\
%
%
&\Rey\left(u_r\frac{\partial u_\theta}{\partial r}+\frac{u_\theta}{r} \frac{\partial u_\theta}{\partial \theta}+\frac{u_r u_\theta}{r}+u_z\frac{\partial u_\theta}{\partial z}\right)=-\frac{1}{r}\frac{\partial p}{\partial r}+\frac{\epsilon^2}{r}\frac{\partial}{\partial r}\left(r\frac{\partial u_\theta}{\partial r}\right)-\epsilon^2\frac{u_\theta}{r^2}+\frac{\epsilon^2}{r^2}\frac{\partial^2 u_\theta}{\partial \theta^2}+\frac{2\epsilon^2}{r^2}\frac{\partial u_r}{\partial \theta}+\frac{\partial^2 u_\theta}{\partial z^2},\\
%
%
&\Rey\epsilon^2\left(u_r\frac{\partial u_z}{\partial r}+\frac{u_\theta}{r}\frac{\partial u_z}{\partial \theta}+u_z\frac{\partial u_z}{\partial z} \right)=-\frac{\partial p}{\partial z}+\frac{\epsilon^4}{r}\frac{\partial}{\partial r}\left(r\frac{\partial u_z}{\partial r}\right)+\frac{\epsilon^4}{r^2}\frac{\partial^2 u_z}{\partial \theta^2}+\epsilon^2\frac{\partial^2 u_z}{\partial z^2},
\end{align}
\end{subequations}
where the Reynolds number is defined as $\Rey=\rho\Omega h_0^2/\mu$, and the gap aspect ratio is $\epsilon=h_0/R$. 
Note that the specific solution to the axisymmetric case with perfectly parallel plates where $\phi = 0$ and small gap heights (i.e. $\epsilon\ll1$) has been derived by \citet{savins1970radial} and is given as follows,
\begin{subequations}\label{eq:axisymmetric}
\begin{align}
&u_{r,\text{axi}}(r,z)=-\frac{1}{12}r\Rey\,z(z-1)\left(-\frac{4}{5}+z+z^2\right)+\mathcal{O}\left(\epsilon^4\right),\\
&u_{\theta,\text{axi}}(r,z)=rz - \frac{r\Rey^2z}{6300}\left(8+z^3\left(35-63z+20z^3\right)\right) + \mathcal{O}\left(\epsilon^4\right),\\
&u_{z,\text{axi}}(z)=\frac{1}{30}\Rey\,z^2(z-1)^2(2+z)+\mathcal{O}\left(\epsilon^4\right),\\
&p_\text{axi}(r)=\frac{3r^2}{20}\Rey + \frac{1}{30}\Rey\,\epsilon^2 z\left(4-9z+5z^3\right) + \mathcal{O}\left(\epsilon^4\right).
\end{align}
\end{subequations}
This study is concerned with the limit of small gap heights, $h_0/R \ll 1$, where the effect of misalignment is most likely to be significant \citep{ault2023viscosity}. Thus, we consider the limit of $\epsilon\ll 1$. Additionally, typical rotation speeds used in rotational rheometers at small gap heights are less than 500 s$^{-1}$ \citep{cardinaels2019quantifying}. Together, these allow us to consider the limit $\Rey \ll 1$, and we neglect inertial effects. With these two assumptions, the resulting leading-order solution for the axisymmetric case with perfectly parallel plates reduces to $u_{\theta} =rz$. 

To derive a theoretical solution for the misaligned case with a nonzero tilt angle $\phi$, we can use the above result in the limit of $\Rey\ll 1$ and consider a perturbation expansion in powers of $\phi/\epsilon$ about the axisymmetric solution. That is, we seek solutions according to
\begin{subequations}\label{eq:expansion}
\begin{gather}
u_r(r,\theta,z)=\left(\frac{\phi}{\epsilon}\right)u_{r,1}(r,\theta,z)+\left(\frac{\phi}{\epsilon}\right)^2u_{r,2}(r,\theta,z)+\hdots,\\
u_\theta(r,\theta,z)=rz+\left(\frac{\phi}{\epsilon}\right)u_{\theta,1}(r,\theta,z)+\left(\frac{\phi}{\epsilon}\right)^2u_{\theta,2}(r,\theta,z)+\hdots,\\
u_z(r,\theta,z)=\left(\frac{\phi}{\epsilon}\right)u_{z,1}(r,\theta,z)+\left(\frac{\phi}{\epsilon}\right)^2u_{z,2}(r,\theta,z)+\hdots,\\
p(r,\theta)=\left(\frac{\phi}{\epsilon}\right)p_1(r,\theta)+\left(\frac{\phi}{\epsilon}\right)^2p_2(r,\theta)+\hdots.
\end{gather}
\end{subequations}
Note that for small tilt angles, the tilt angle $\phi$ can range from 0 to a maximum of $\epsilon$, so that $\phi/\epsilon$ ranges from 0 to 1. Small values of $\phi/\epsilon$ correspond to small plate deflections, and $\phi/\epsilon=1$ corresponds to when the plates come in contact. In the limits $\epsilon\ll 1$ and $Re\ll 1$, the nondimensional governing equations (\ref{eq:variableNS_cylindrical}) simplify to
\begin{subequations}\label{eq:governing}
\begin{align}
&0=-\frac{\partial p}{\partial r} +\frac{\partial^2u_r}{\partial {z}^2},\\
&0=-\frac{1}{r}\frac{\partial p}{\partial\theta}+\frac{\partial^2u_\theta}{\partial {z}^2},\\
&0=\frac{\partial p}{\partial z}.
\end{align}
\end{subequations}

The boundary conditions for the system are given by $\mathbf{u}=0$ at $z=0$ on the lower plate of the rheometer, and 
\begin{subequations}\label{eq:bc}
    \begin{gather}
        u_r = r\cos\theta\sin\theta\sin\phi\tan\phi,\\
        u_\theta=r(\cos^2\theta\sec\phi+\cos\phi\sin^2\theta),\\
        u_z=-r\epsilon^{-1} \sin \theta \sin \phi,
    \end{gather}
\end{subequations}
at $z=h(r,\theta,\phi)=1+\phi\epsilon^{-1} r \cos\theta$ on the upper plate of the rheometer. 

For the generalized case of Newtonian, single-phase fluid dynamics in a rheometer with misaligned plates, the perturbation expansion (\ref{eq:expansion}) can be substituted into the governing equations (\ref{eq:governing}), and the boundary condition equations (\ref{eq:bc}) can be applied to yield solutions for the higher-order velocity and pressure terms given by
\begin{subequations}\label{eq:solution}
\begin{align}
    u_{r,1}(r,\theta,z)&=-\frac{3}{8}\left(-1+3r^2\right)(-1+z)z\sin\theta ,\\
    u_{r,2}(r,\theta,z)&=\frac{3}{16}rz\left(4-5z+r^2(-7+10z)\right)\sin(2\theta) ,\\
    u_{\theta,1}(r,\theta,z)&=-\frac{1}{8}z\left(3-3z+r^2(5+3z)\right)\cos\theta ,\\
    u_{\theta,2}(r,\theta,z)&=\frac{1}{16}rz\left(-3+11r^2+\left(12-15z+r^2(-4+15z)\right)\cos(2\theta)\right) ,\\
    u_{z,1}(r,\theta,z)&=r(-2+z)z^2\sin\theta ,\\
    u_{z,2}(r,\theta,z)&=\frac{1}{8}r^2(19-15z)z^2\sin(2\theta) ,\\
    p_1(r,\theta)&= -\frac{3}{4}r\left(-1+r^2\right)\sin\theta,\\
    p_2(r,\theta)&= \frac{15}{16}r^2\left(r^2-1\right)\sin(2\theta).
\end{align}
\end{subequations}
With the theoretical solutions for pressure and velocity field, we can solve for the forces and moments acting on the top plate of the rheometer due to pressure and viscous effects. The dimensionless vertical force (nondimensionalized by $\mu\Omega R^4/h^2_0$) due to pressure can be calculated by
\begin{equation}\label{eq:pressure}
    F_{\textrm{pressure},z} = \int^{2\pi}_0 \int^1_0 pr drd\theta.
\end{equation}
The vertical pressure force that arises directly from the misalignment of the plates, when calculated using the pressure solution of equation (\ref{eq:solution}), yields a value of 0. This is because of the symmetry of the misalignment pressure. The increased pressure on the converging side of the gap is balanced by reduced pressure on the diverging side of the gap. For completeness, inertial effects, which are assumed to be small here, do give rise to a vertical pressure force. This value can be found by considering the previous solution given in equations (\ref{eq:axisymmetric}), which was derived using the full governing equations (\ref{eq:variableNS_cylindrical}) and still retains the effects of inertia. Using the pressure equation from this solution, the vertical force due to pressure can be calculated with equation (\ref{eq:pressure}), and is found to be $F_{\textrm{pressure},z} = \frac{6\pi\Rey}{80}$. As expected, in the limit of small $\Rey$, this pressure force is close to zero. 

The force (nondimensionalized by $\mu\Omega R^3/h_0$) that arises due to viscous effects can be obtained by
\begin{equation}
    \boldsymbol{F}_{\textrm{viscous}} = \int^{2\pi}_0 \int^1_0 \frac{\partial \boldsymbol{U}}{\partial z} rdrd\theta = 0\boldsymbol{i}-\frac{\pi(\phi/\epsilon)}{4}\boldsymbol{j}+0\boldsymbol{k}.
\end{equation}
Here, we converted $\boldsymbol{U}$ from cylindrical coordinates to Cartesian coordinates by using $u_x = u_r \cos\theta - u_\theta\sin\theta$ and $u_y = u_r\sin\theta+u_\theta\cos\theta$. Next, the dominant pressure moment (nondimensionalized by $\mu\Omega R^5/h^2_0$) due to the primary flow and misalignment can be calculated by integrating $dM_x = dFy$, where $dF = pdA$, and is found to be
\begin{equation}
    M_{\textrm{pressure},x} = \int^{2\pi}_0 \int^1_0 pyr drd\theta = \int^{2\pi}_0 \int^1_0 pr^2\sin\theta drd\theta = \frac{\pi(\phi/\epsilon)}{16}.
\end{equation}
The dominant viscous moment (nondimensionalized by $\mu\Omega R^4/h_0$) due to misalignment, which acts along the $z$-direction, is found to be
\begin{equation}\label{vis_mom}
    M_{\textrm{viscous},z} = \int^{2\pi}_0 \int^1_0 \frac{\partial u_\theta}{\partial z} \Bigr|_{z=1} r^2 drd\theta = \frac{1}{96}\pi\left(48+13(\phi/\epsilon)^2\right).
\end{equation}
The theoretical solutions have shown that in the limit of $\Rey\ll 1$, the pressure force in the $z$-direction is very small, and the dominant force is the viscous force in the $y$-direction. Additionally, the theoretical solution also predicts the role of misalignment on the viscous moment in the $z$-direction, which is due to the rotation of the upper plate and the viscosity of the fluid. Using the theoretical results of forces and moments, we can estimate the viscosity measurement of a misaligned rheometer, the results of which will be shown in the following sections. 

Using the viscous moment predicted by Equation \ref{vis_mom}, we can estimate the error in viscosity measurements in a misaligned parallel-plate rheometer. Thus, the true viscosity can be recovered from the measured viscosity using the equation,
\begin{equation}
    \frac{\mu_\textrm{measured}}{\mu_\textrm{true}} = \frac{48+13(\phi/\epsilon)^2}{48}.
\end{equation}
This relation shows that as we increase the misalignment in the rheometer, there is an increase in error in the viscosity measurement by the rheometer. In the following sections, we will compare theoretical predictions with numerical simulations.

\section{Numerical simulation}\label{numerical}
To validate the theoretical derivation above, we performed numerical simulations for the parallel-plate rheometer with misalignment. Three-dimensional numerical OpenFOAM simulations were used to solve the Newtonian, single-phase fluid dynamics within the parallel-plate rheometer. During the simulation, the bottom plate of the rheometer was flat and held stationary, while the upper plate was tilted at an angle $\phi$ and was set to rotate at an angular velocity of $\Omega$. We performed simulations for tilt angles of $\phi/\epsilon = 0.01, 0.02,0.05,0.08,0.1,0.2$, and 0.5, with $\epsilon$ fixed at 0.1 for all of the cases. The 3D simulations fully resolved the velocity and pressure field within the gap of the rheometer. The pressure force, viscous force, viscous moment, and pressure moment acting on the top plate were calculated from the results of the simulations.  The computational setup and convergence study are detailed in Appendix \ref{Appendix:A}.

\begin{figure}
    \centering
    \includegraphics[width=\textwidth]{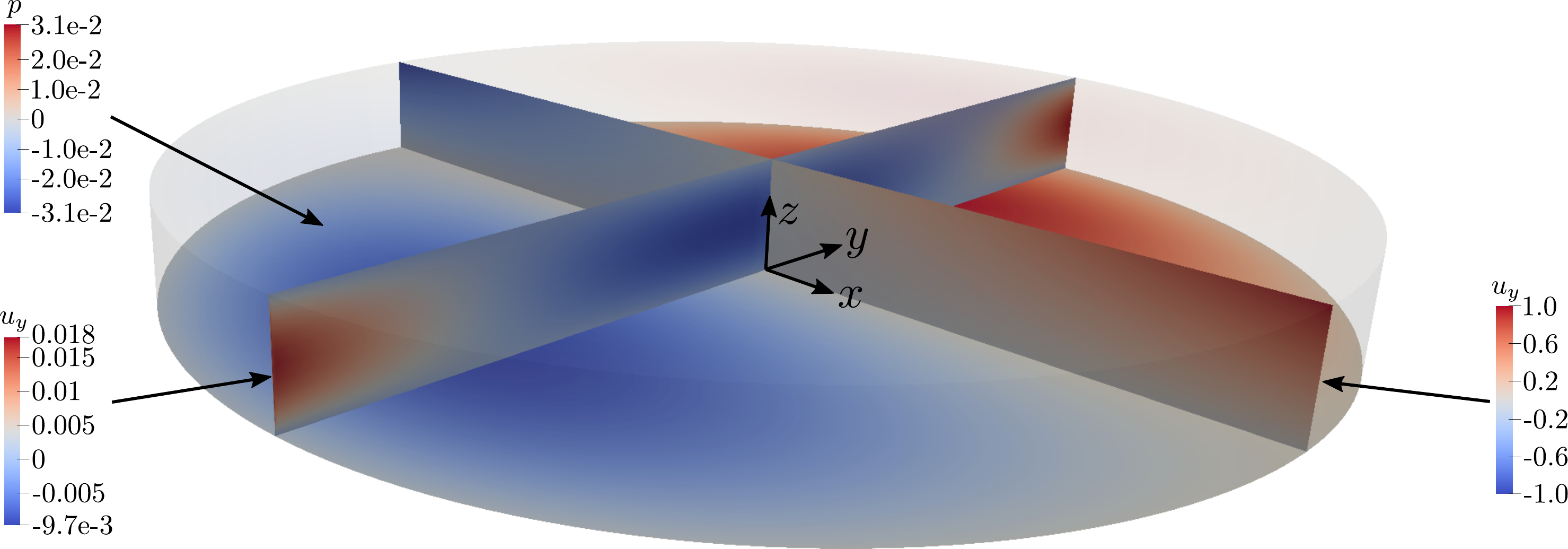}
    \caption{Secondary flow structure of the misaligned parallel-plate rheometer for the case of $\epsilon=0.01$, $\phi/\epsilon=0.1$, and $\Rey=10^{-4}$. The cross-section plane shows the recirculation arises from the misalignment.}
    \label{fig:full3D}
\end{figure}

A typical result for the misaligned case of a parallel-plate rheometer is shown in Figure \ref{fig:full3D}, which shows the pressure and velocity components due to misalignment. On the $x$-$z$ plane, we observe the primary azimuthal velocity (presented as $u_y$) driven by the rotating upper plate. As we move away from the center of the rheometer and move up in the $z$-direction, we observe an increase in velocity magnitude that is due to the azimuthal velocity of the rotating top plate. At $x=-1$, the gap between the plates is at a minimum, resulting in a pressure gradient inside the rheometer. Larger pressure is observed for $y>0$ due to the fluid being squeezed by the gap. Finally, the resulting non-axisymmetric pressure drives the flow in the negative $y$-direction along the center of the system, which can be observed in the $y$-$z$ plane velocity component.

\begin{figure}
    \centering
    \includegraphics[width=\textwidth]{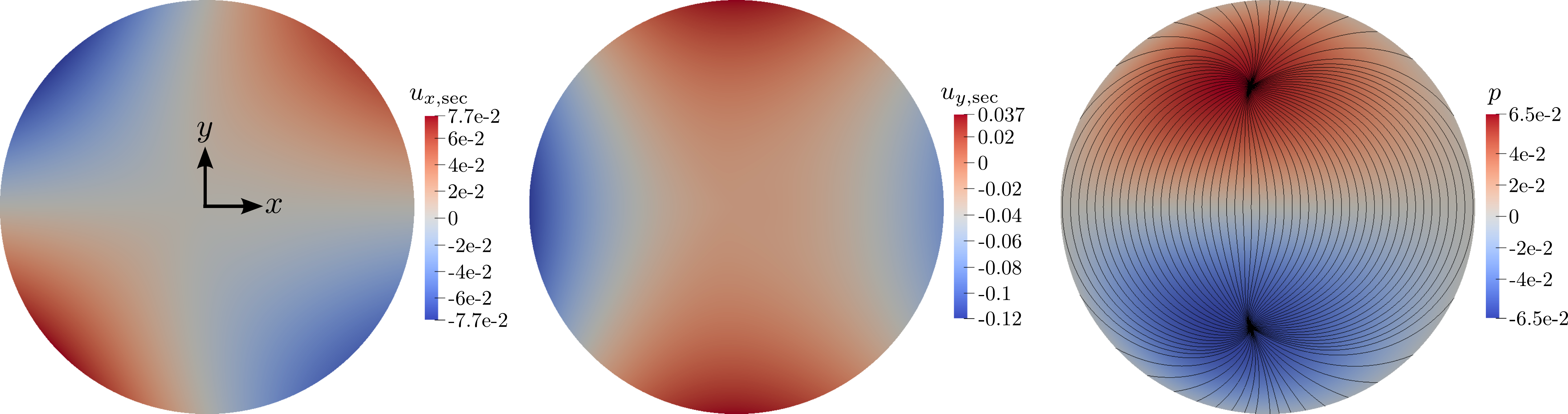}
    \caption{Numerical simulation results for the secondary $x$ and $y$ velocity components and pressure, taken across the cross-section through the $z=1/2$ plane in a misaligned rheometer with $\epsilon=0.01$, $\phi/\epsilon=0.2$, and $\Rey = 10^{-4}$. The left panel shows the secondary velocity in the $x$-direction, the middle panel shows the secondary velocity in the $y$-direction, and the right panel shows the pressure on the cross-section along with streamlines computed from the secondary velocity components.}
    \label{fig:secondary}
\end{figure}

To better visualize the secondary flow behaviors inside the misaligned rheometer, we present the secondary flow patterns in the cross-section on the $z=1/2$ plane in Figure \ref{fig:secondary}. Here, the numerical simulation was conducted for the case of $\epsilon=0.01$, $\phi/\epsilon=0.2$, and $\Rey=10^{-4}$, and the $x$ component of velocity, the $y$ component of velocity, and the pressure are plotted on the cross-section.
The pressure is plotted along with the secondary flow streamlines on the cross-section; the primary flow  $u_\theta=rz$ is subtracted, and the streamlines are then calculated to give a representation of the secondary flow patterns. The secondary streamlines all flow away from the high pressure region on top and towards the low pressure region on the bottom. The order of magnitude of the secondary velocities ranges up to around 0.1, which is the same order of magnitude as $\phi/\epsilon$. The $O(\Rey)$ secondary flow typically seen in a parallel-plate rheometer is negligible in this case.

\section{Results}\label{result}
Previously, we developed theoretical predictions for the velocity and pressure profiles in the non-axisymmetric parallel-plate rheometer with misaligned plates. We also performed 3D numerical simulations of the Newtonian, single-phase Navier-Stokes equations to visualize and understand the fluid dynamics in the system. In this section, we will compare the theoretical results with numerical simulations and provide a more detailed analysis of the physics inside the misaligned rheometer.

\begin{figure}
    \centering
    \includegraphics[width=\textwidth]{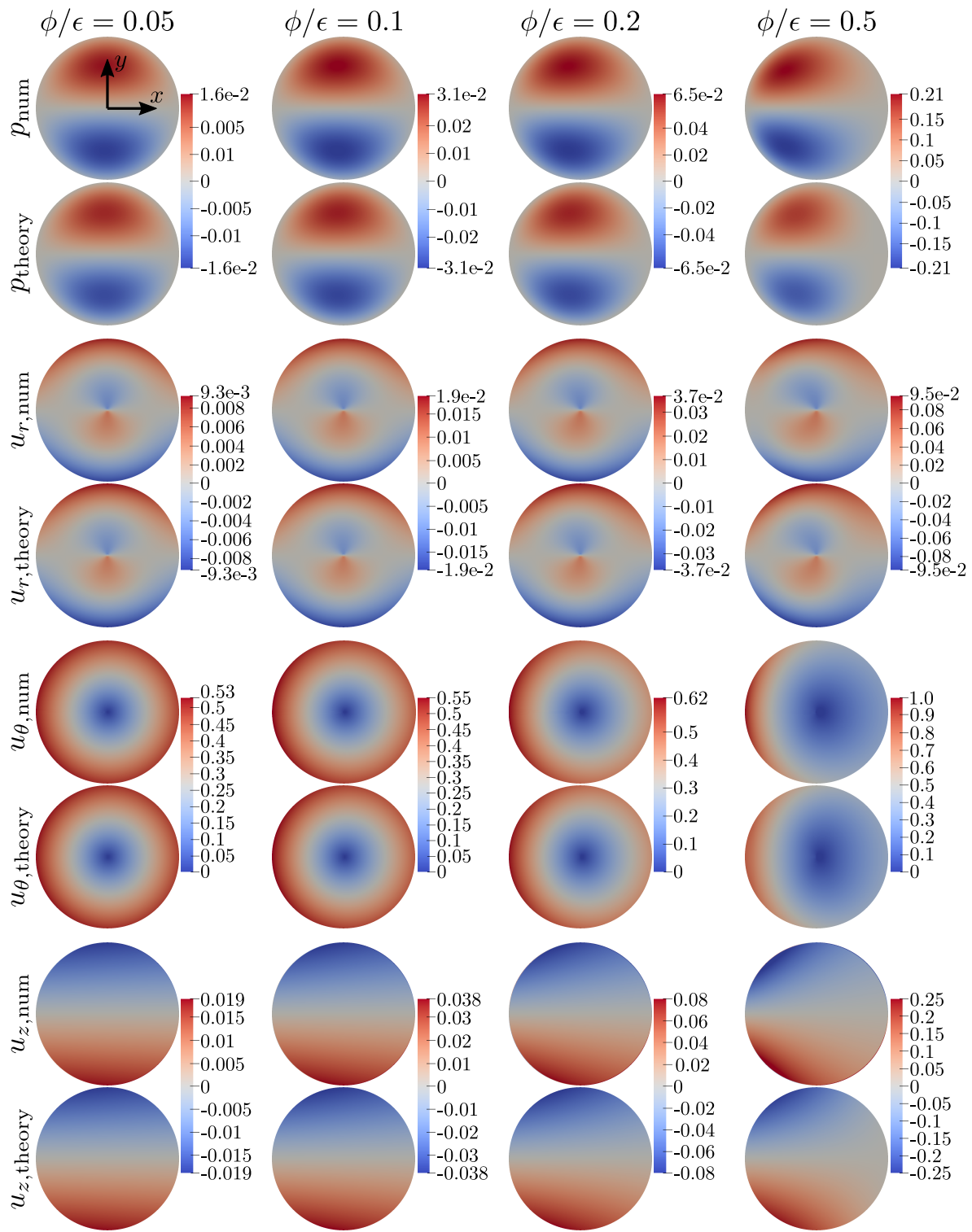}
    \caption{Numerical and theoretical velocity/pressure profiles in a parallel-plate rheometer with misalignment for $\phi/\epsilon$ values of 0.05, 0.1, 0.2, and 0.5. All results are plotted at $z=1/2$, with $\epsilon=0.01$ and $\Rey = 10^{-4}$. A non-axisymmetric pressure gradient builds in response to the fluid being forced into a narrowing and then expanding gap. For small $\phi/\epsilon$, the velocities and pressures are symmetric about both the $x$- and $y$- axes, but the symmetry about the $y$-axis breaks for larger values of $\phi/\epsilon$.}
    \label{fig:crosssection}
\end{figure}

Figure \ref{fig:crosssection} shows plots of the theoretical predictions of the pressure profiles $p$ and the velocity components $u_r$, $u_\theta$, $u_z$, for $\phi/\epsilon$ values of 0.05, 0.1, 0.2, and 0.5. The results are plotted on the cross section of $z=1/2$, with $\epsilon=0.01$ and $\Rey = 10^{-4}$. The plots for the pressure profiles show that the pressure varies like the $\sin\theta$ function. The rotating upper-plate drives the flow, and the pressure results from the flow being forced into a narrowing or diverging gap. Near the center of the rheometer, where the fluid has relatively low velocities, this emergent pressure gradient drives a flow in the negative $y$-direction, which can be seen in the results for $u_r$. As $\phi/\epsilon$ increases, the symmetry about the $y$-axis begins to break as the $O(\phi^2/\epsilon^2)$ pressure correction grows, which has a $\sin(2\theta)$ functional dependence. The plots for $u_{\theta}$ show that velocity is the slowest near the center. The $u_z$ plots show that the normal velocity inside the rheometer plates is symmetric along the $y$-axis. As $\phi/\epsilon$ increases, the dominant pressure and velocity gradients shift to the left side of the rheometer, where the distance between the upper and lower plates is the smallest. All of these results support our previously proposed theory for the velocity and pressure profiles of the misaligned rheometer.

\begin{figure}
    \centering
    \includegraphics[width=\textwidth]{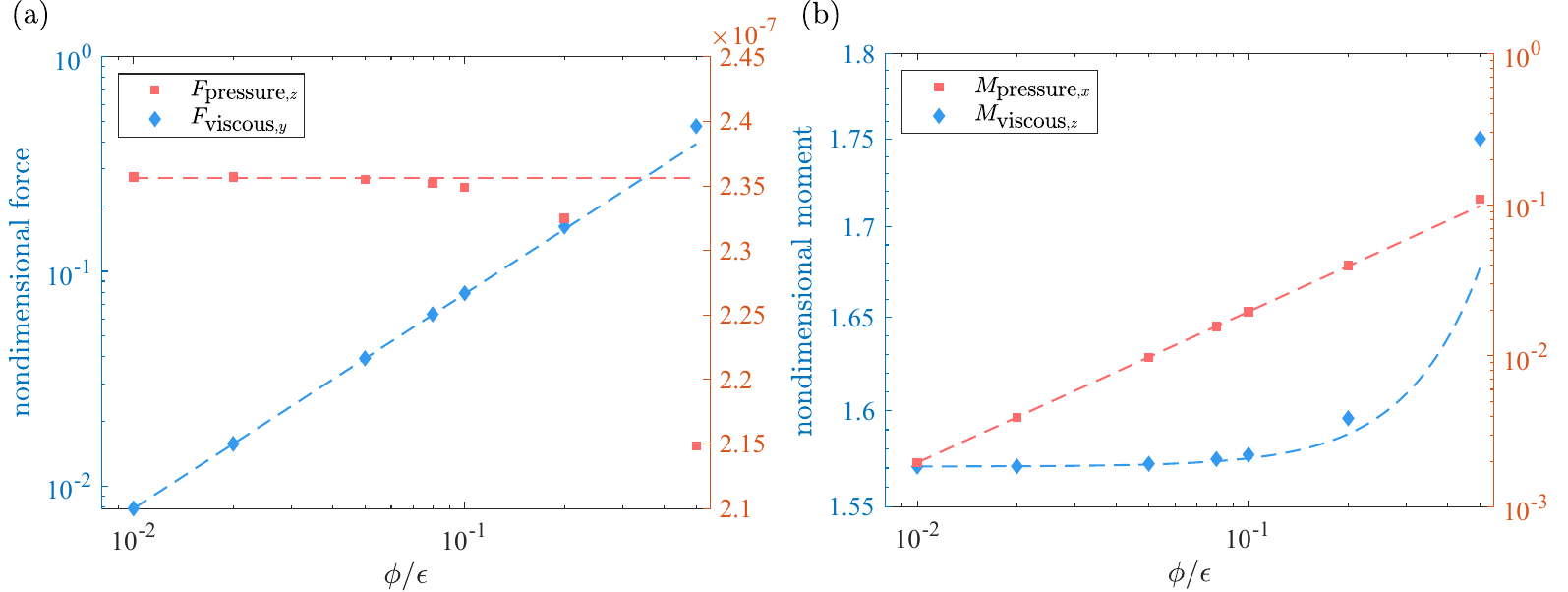}
    \caption{Theoretical predictions and numerical simulations of the nondimensional forces and moments on the upper plate of a parallel-plate rheometer with misalignment as a function of $\phi/\epsilon$ with $\epsilon=0.01$. The dashed-lines correspond to the theoretical predictions, and the symbols correspond to the results from numerical simulations.}
    \label{fig:force_moment}
\end{figure}

\begin{figure}
    \centering
    \includegraphics[width=0.5\textwidth]{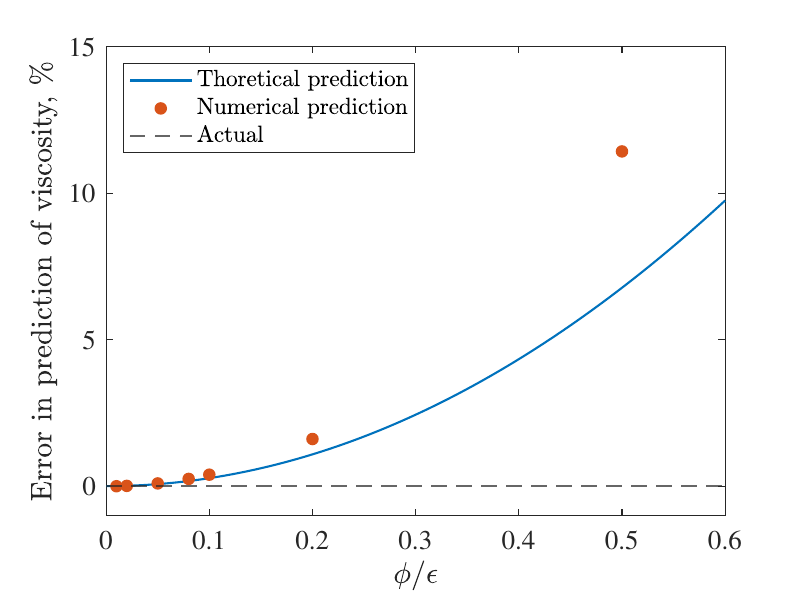}
    \caption{Percentage error in prediction of viscosity by a parallel plate rheometer with misalignment as a function of $\phi/\epsilon$. The true viscosity is represented by the dashed-line at 0\%. The solid-line corresponds to the theoretical estimate with misalignment derived using Equation \ref{vis_mom}. The symbols correspond to the results from numerical simulations. As the misalignment increases, the error in the prediction of viscosity by the rheometer increases.}
    \label{fig:pred_vis}
\end{figure}

In addition to the pressure and velocity profiles, the forces and moments acting on the upper plate of the rheometer for each case were also calculated from the equations derived in Section \ref{theory}. From the theoretical predictions, there is a pressure force acting in the $z$-direction, a viscous force acting in the $y$-direction, a pressure moment acting in the $x$-direction, and a viscous moment acting in the $z$-direction as previously stated, which we also calculated numerically. Figure \ref{fig:force_moment} shows both the theoretical predictions and the numerical simulation results of the nondimensional forces and moments acting on the upper plate of the rheometer as functions of $\phi/\epsilon$, with $\epsilon=0.01$ and $\Rey=10^{-4}$. The dashed lines are the theoretical predictions, and the solid points are the numerical results. The numerical results agree well with the theoretical predictions.
The pressure force shown in Figure \ref{fig:force_moment}(a) is small in magnitude due to the low Reynolds number and is independent of $\phi/\epsilon$, in contrast to the viscous force. Here, inertia is negligible, and the viscous force is the dominant force in the system, and increases with $\phi/\epsilon$. The viscous moment shown in Figure \ref{fig:force_moment}(b) is smaller than the pressure moment as seen in the same figure. As $\phi/\epsilon$ increases, both moments increase, but the viscous moment increases at a higher rate when $\phi/\epsilon$ is large. From the theoretical predictions and numerical simulations, we see that as the misalignment $\phi/\epsilon$ increases, the effect on the force and moment also increases. Since viscosity measurements are based on the measured forces and moments acting on the rheometer by the fluid, misalignment in a parallel plate rheometer, especially at large tilt angles $\phi/\epsilon$, can significantly impact the measurement of the viscosity of the fluid. Using the theoretical prediction derived from Section \ref{theory}, we can predict the theoretical viscosity measured when using a misaligned rheometer and compare it with the true viscosity as well as numerical simulation results. Figure \ref{fig:pred_vis} shows the percentage error in viscosity prediction by a parallel plate rheometer with misalignment as a function of $\phi/\epsilon$. The solid-line corresponds to the theoretical estimate with misalignment derived using Equation \ref{vis_mom}. The symbols correspond to the estimates from numerical simulations. The percentage error is calculated by comparing the estimated viscosity with the true viscosity of the fluid.  As shown in Figure \ref{fig:pred_vis}, as the misalignment increases, the rheometer sees an increase in error by a greater magnitude. The error approaches and exceeds 10\% as $\phi/\epsilon$ exceeds 0.5. In practice, to notice a misaligned rheometer, one should perform at least two rounds of viscosity measurements of the same sample at different gap heights. Under low-Reynolds-number conditions, if decreasing the gap height causes an increase in moments and forces as well as changes in viscosity prediction, there is a possibility that the plates of the rheometer are misaligned sufficiently to impact the measurements.

\section{Conclusions}\label{conclusion}
In this study, we considered the Newtonian fluid dynamics in a misaligned parallel-plate rheometer. This work was motivated by the recent paper \cite{ault2023viscosity} that considered the measurement of the viscosity of glycerol, where it was found that the misalignment of the plates of the rheometer induces higher absorption of water vapor from the atmosphere due to modified secondary flows and thus can lead to inaccurate measurements of viscosity. Here, we derived the theoretical predictions for pressure and velocity in the rheometer by modeling the fluid dynamics in the misaligned geometry. The Navier-Stokes equations were solved using perturbation expansions, and the pressure and velocity profiles of the system were obtained. We also solved for the forces and moments acting on the upper plate of the parallel-plate rheometer. We predicted the viscosity measurement under misaligned conditions from the theoretical force and moment calculations. Full 3D numerical simulations were performed using OpenFOAM to validate our theoretical predictions, and the theoretical predictions show good agreement with the numerical simulations. From both the theoretical predictions and the numerical simulations, we observed that the dominant forces and moments in the misaligned rheometer are the viscous force in the $y$-direction, the pressure moment in the $x$-direction, and the viscous moment in the $z$-direction. Based on these results, the misalignment of plates in a parallel-plate rheometer can contribute to the inaccurate reading of viscosity measurements, as when misalignment increases, the magnitudes of the forces and moments acting on the upper plate of the rheometer increases, making an increase of error in the prediction of viscosity by the rheometer. This work provides a simple approach to correct for misalignment errors in a parallel-plate rheometer when working with small gap heights.

\appendix

\section{Numerical details}\label{Appendix:A}
In this section, we will discuss the detail of the numerical simulations performed, as well as the numerical convergence tests performed. The numerical simulations were performed using OpenFOAM with the simpleFOAM solver \citep{weller1998tensorial}, which uses the ``semi-implicit method for pressure-linked equations'' (SIMPLE) algorithm \citep{ferziger2002computational}. The SIMPLE algorithm is specifically designed for steady-state problems and does not need to fully resolve the pressure-velocity coupling. The OpenFOAM simulations were second-order accurate spatially, and the relaxation factors for velocity and pressure were set to 0.7 and 0.3, respectively. The solver was iterated until the tolerance of $10^{-10}$ was met for both pressure and velocity for each case, and the fluid dynamics within the rheometer converged to the steady state. 

\begin{figure}
    \centering
    \includegraphics[width=0.8\textwidth]{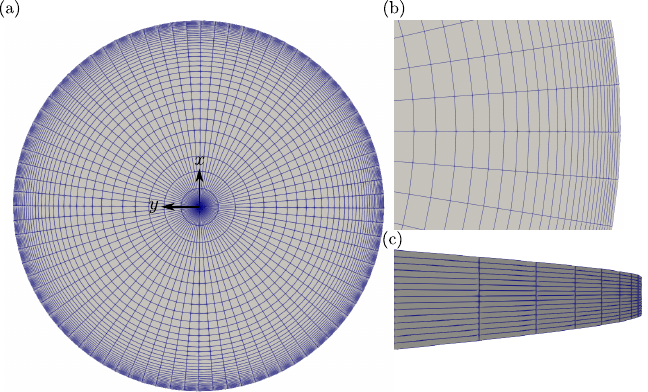}
    \caption{Sample simulation domain of a parallel-plate rheometer. (a) Top view of the simulation domain. (b) Zoomed-in top view showing the local refinement near the edge of the rheometer. (c) Side view of the rheometer simulation domain. }
    \label{fig:mesh}
\end{figure}

The mesh grid in the simulation was constructed by using the BlockMesh utility in the OpenFOAM library. The bottom plate was held stationary while an angular velocity, was imposed on the upper plate of the rheometer. The upper plate was tilted at an angle $\phi$ in the misaligned case. Figure \ref{fig:mesh} shows the simulation domain and mesh grid. Figure \ref{fig:mesh}(a) shows the top view of the simulation domain. The outer boundary condition was set to be a slip boundary condition, and a local refinement near the edge was imposed to accurately resolve the fluid dynamics near the boundary as shown in Figure \ref{fig:mesh}(b). In the depth direction, a uniform cell thickness was used, as shown in Figure \ref{fig:mesh}(c).

\begin{figure}
    \centering
    \includegraphics[width=0.7\textwidth]{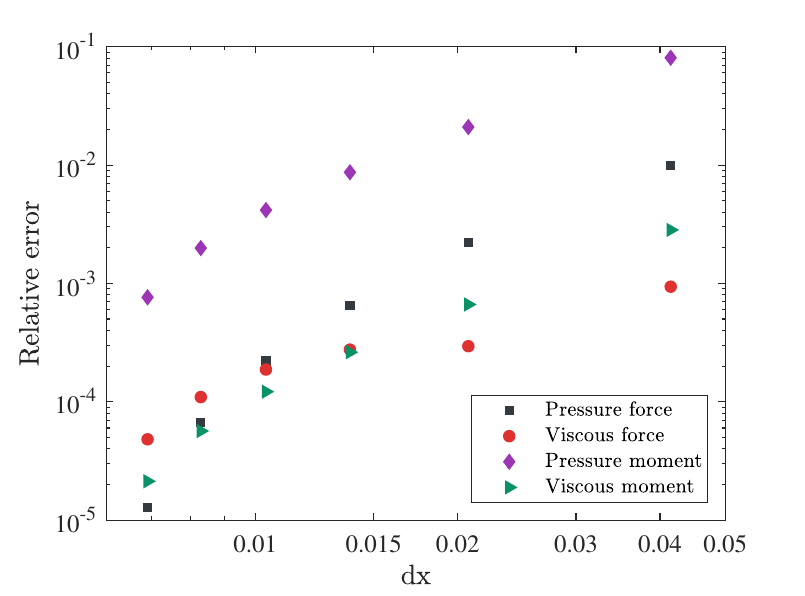}
    \caption{The relative error of forces and moments as functions of grid size with $\phi/\epsilon=0.01$. The forces and moments are calculated by both theoretical prediction and numerical simulations. The relative errors are calculated by comparing numerical simulations with the theoretical predictions. As dx decreases, the relative error decreases, and all are below $0.1\%$ for the finest grid.}
    \label{fig:conver}
\end{figure}

The simulations were performed for $\phi/\epsilon = 0.01, 0.02,0.05,0.08,0.1,0.2$, and 0.5. The value of $\epsilon$ was fixed at 0.01, and the radius of the rheometer was fixed at 1. To confirm numerical convergence, we calculate the relative error of the numerical forces and moments with $\phi/\epsilon=0.01$ for various grid sizes. Figure \ref{fig:conver} shows the convergence study of the numerical simulation using the relative error of forces and moments by comparing theory and simulation. As shown in Figure \ref{fig:conver}, the relative error decreases as dx decreases. The relative errors are all less than $10^{-3}$ with the smallest dx, providing accurate predictions of the forces and moments to within $0.1\%$. 

\bibliography{reference.bib}

\begin{thebibliography}{34}
\providecommand{\natexlab}[1]{#1}
\providecommand{\url}[1]{\texttt{#1}}
\expandafter\ifx\csname urlstyle\endcsname\relax
  \providecommand{\doi}[1]{doi: #1}\else
  \providecommand{\doi}{doi: \begingroup \urlstyle{rm}\Url}\fi

\bibitem[Ahmed(2018)]{ahmed2018advances}
J.~Ahmed.
\newblock Advances in rheological measurements of food products.
\newblock \emph{Curr. Opin. Food Sci.}, 23:\penalty0 127--132, 2018.

\bibitem[Andablo-Reyes et~al.(2010)Andablo-Reyes, Hidalgo-{\'A}lvarez, and
  De~Vicente]{andablo2010method}
E.~Andablo-Reyes, R.~Hidalgo-{\'A}lvarez, and J.~De~Vicente.
\newblock A method for the estimation of the film thickness and plate tilt
  angle in thin film misaligned plate--plate rheometry.
\newblock \emph{J. Non-Newtonian Fluid Mech.}, 165\penalty0 (19-20):\penalty0
  1419--1421, 2010.

\bibitem[Andablo-Reyes et~al.(2011)Andablo-Reyes, Vicente, and
  Hidalgo-Alvarez]{andablo2011nonparallelism}
E.~Andablo-Reyes, J.~Vicente, and R.~Hidalgo-Alvarez.
\newblock On the nonparallelism effect in thin film plate--plate rheometry.
\newblock \emph{J. Rheol.}, 55\penalty0 (5):\penalty0 981--986, 2011.

\bibitem[Ault et~al.(2023)Ault, Shin, Garcia, Perazzo, and
  Stone]{ault2023viscosity}
J.~T. Ault, S.~Shin, A.~Garcia, A.~Perazzo, and H.~A. Stone.
\newblock Viscosity measurements of glycerol in a parallel-plate rheometer
  exposed to atmosphere.
\newblock \emph{J. Fluid Mech.}, 968:\penalty0 A2, 2023.

\bibitem[Cardinaels et~al.(2019)Cardinaels, Reddy, and
  Clasen]{cardinaels2019quantifying}
R.~Cardinaels, N.K. Reddy, and C.~Clasen.
\newblock Quantifying the errors due to overfilling for newtonian fluids in
  rotational rheometry.
\newblock \emph{Rheol. Acta}, 58:\penalty0 525--538, 2019.

\bibitem[Cardoso et~al.(2015)Cardoso, Fujii, Pileggi, and
  Chaouche]{cardoso2015parallel}
F.A. Cardoso, A.L. Fujii, R.G. Pileggi, and M.~Chaouche.
\newblock Parallel-plate rotational rheometry of cement paste: Influence of the
  squeeze velocity during gap positioning.
\newblock \emph{Cem. Concr. Res.}, 75:\penalty0 66--74, 2015.

\bibitem[Carotenuto and Minale(2013)]{carotenuto2013use}
C.~Carotenuto and M.~Minale.
\newblock On the use of rough geometries in rheometry.
\newblock \emph{J. Non-Newtonian Fluid Mech.}, 198:\penalty0 39--47, 2013.

\bibitem[Choi and Park(2010)]{choi2010microfluidic}
S.Y. Choi and J.K. Park.
\newblock Microfluidic rheometer for characterization of protein unfolding and
  aggregation in microflows.
\newblock \emph{Small}, 6\penalty0 (12):\penalty0 1306--1310, 2010.

\bibitem[Clasen(2013)]{clasen2013self}
C.~Clasen.
\newblock A self-aligning parallel plate (sapp) fixture for tribology and high
  shear rheometry.
\newblock \emph{Rheol. Acta}, 52\penalty0 (3):\penalty0 191--200, 2013.

\bibitem[Connelly and Greener(1985)]{connelly1985high}
R.W. Connelly and J.~Greener.
\newblock High-shear viscometry with a rotational parallel-disk device.
\newblock \emph{J. Rheol.}, 29\penalty0 (2):\penalty0 209--226, 1985.

\bibitem[Coussot(2005)]{coussot2005rheometry}
Philippe Coussot.
\newblock \emph{Rheometry of pastes, suspensions, and granular materials:
  applications in industry and environment}.
\newblock John Wiley \& Sons, 2005.

\bibitem[Covas et~al.(2000)Covas, N{\'o}brega, and Maia]{covas2000rheological}
J.A. Covas, J.M. N{\'o}brega, and J.M. Maia.
\newblock Rheological measurements along an extruder with an on-line capillary
  rheometer.
\newblock \emph{Polym. Test.}, 19\penalty0 (2):\penalty0 165--176, 2000.

\bibitem[Davies and Stokes(2005)]{davies2005gap}
G.A. Davies and J.R. Stokes.
\newblock On the gap error in parallel plate rheometry that arises from the
  presence of air when zeroing the gap.
\newblock \emph{J. Rheol.}, 49\penalty0 (4):\penalty0 919--922, 2005.

\bibitem[Davies and Stokes(2008)]{davies2008thin}
G.A. Davies and J.R. Stokes.
\newblock Thin film and high shear rheology of multiphase complex fluids.
\newblock \emph{J. Non-Newtonian Fluid Mech.}, 148\penalty0 (1-3):\penalty0
  73--87, 2008.

\bibitem[Desprat et~al.(2006)Desprat, Guiroy, and
  Asnacios]{desprat2006microplates}
N.~Desprat, A.~Guiroy, and A.~Asnacios.
\newblock Microplates-based rheometer for a single living cell.
\newblock \emph{Rev. Sci. Instrum.}, 77\penalty0 (5), 2006.

\bibitem[Ferraris and Martys(2003)]{ferraris2003relating}
C.F. Ferraris and N.S. Martys.
\newblock Relating fresh concrete viscosity measurements from different
  rheometers.
\newblock \emph{J. Res. Natl. Inst. Stand. Technol.}, 108\penalty0
  (3):\penalty0 229, 2003.

\bibitem[Ferziger et~al.(2002)Ferziger, Peri{\'c}, and
  Street]{ferziger2002computational}
J.~H. Ferziger, M.~Peri{\'c}, and R.L. Street.
\newblock \emph{Computational methods for fluid dynamics}, volume~3.
\newblock Springer, 2002.

\bibitem[Gallegos and Franco(1999)]{gallegos1999rheology}
C.~Gallegos and J.M. Franco.
\newblock Rheology of food, cosmetics and pharmaceuticals.
\newblock \emph{Curr. Opin. Colloid Interface Sci.}, 4\penalty0 (4):\penalty0
  288--293, 1999.

\bibitem[Giacomin et~al.(1989)Giacomin, Samurkas, and Dealy]{giacomin1989novel}
A.J. Giacomin, T.~Samurkas, and J.M. Dealy.
\newblock A novel sliding plate rheometer for molten plastics.
\newblock \emph{Polym. Eng. Sci.}, 29\penalty0 (8):\penalty0 499--504, 1989.

\bibitem[Goh et~al.(2010)Goh, Versluis, Appelqvist, and
  Bialek]{goh2010tribological}
S.M. Goh, P.~Versluis, I.A.M. Appelqvist, and L.~Bialek.
\newblock Tribological measurements of foods using a rheometer.
\newblock \emph{Food Res. Int.}, 43\penalty0 (1):\penalty0 183--186, 2010.

\bibitem[Hellstr{\"o}m et~al.(2014)Hellstr{\"o}m, Samaha, Wang, Smits, and
  Hultmark]{hellstrom2014errors}
L.H.O. Hellstr{\"o}m, M.A. Samaha, K.M. Wang, A.J. Smits, and M.~Hultmark.
\newblock Errors in parallel-plate and cone-plate rheometer measurements due to
  sample underfill.
\newblock \emph{Meas. Sci. Technol.}, 26\penalty0 (1):\penalty0 015301, 2014.

\bibitem[Ihm et~al.(2020)Ihm, Lee, Ahn, and Oh]{ihm2020viscosity}
C.~Ihm, D.S. Lee, K.H. Ahn, and J.S. Oh.
\newblock Viscosity measurement of whole blood with parallel plate rheometers.
\newblock \emph{Biochip J.}, 14:\penalty0 179--184, 2020.

\bibitem[Jones et~al.(1997)Jones, Woolfson, and Brown]{jones1997textural}
D.S. Jones, A.D. Woolfson, and A.F. Brown.
\newblock Textural, viscoelastic and mucoadhesive properties of pharmaceutical
  gels composed of cellulose polymers.
\newblock \emph{Int. J. Pharm.}, 151\penalty0 (2):\penalty0 223--233, 1997.

\bibitem[Kramer et~al.(1987)Kramer, Uhl, and Prud'Homme]{kramer1987measurement}
J.~Kramer, J.T. Uhl, and R.K. Prud'Homme.
\newblock Measurement of the viscosity of guar gum solutions to 50,000 $s^{-1}$
  using a parallel plate rheometer.
\newblock \emph{Polym. Eng. Sci.}, 27\penalty0 (8):\penalty0 598--602, 1987.

\bibitem[Macosko(1994)]{macosko1994rheology}
C.W. Macosko.
\newblock \emph{Rheology: principles, measurements, and applications}.
\newblock Wiley-VCH, 1994.

\bibitem[Magnin and Piau(1990)]{magnin1990cone}
A.~Magnin and J.M. Piau.
\newblock Cone-and-plate rheometry of yield stress fluids. study of an aqueous
  gel.
\newblock \emph{J. Non-Newtonian Fluid Mech.}, 36:\penalty0 85--108, 1990.

\bibitem[Malkin and Isayev(2022)]{malkin2022rheology}
A.Y. Malkin and A.I. Isayev.
\newblock \emph{Rheology: concepts, methods, and applications}.
\newblock Elsevier, 2022.

\bibitem[Navaneethan et~al.(2005)Navaneethan, Missaghi, and
  Fassihi]{navaneethan2005application}
C.V. Navaneethan, S.~Missaghi, and R.~Fassihi.
\newblock Application of powder rheometer to determine powder flow properties
  and lubrication efficiency of pharmaceutical particulate systems.
\newblock \emph{AAPS PharmSciTech}, 6:\penalty0 E398--E404, 2005.

\bibitem[Park and Song(2010)]{park2010rheological}
E.K. Park and K.W. Song.
\newblock Rheological evaluation of petroleum jelly as a base material in
  ointment and cream formulations: steady shear flow behavior.
\newblock \emph{Arch. Pharmacal Res.}, 33:\penalty0 141--150, 2010.

\bibitem[Rodr{\'\i}guez-L{\'o}pez et~al.(2013)Rodr{\'\i}guez-L{\'o}pez, Elvira,
  de~Espinosa~Freijo, and De~Vicente]{rodriguez2013using}
J.~Rodr{\'\i}guez-L{\'o}pez, L.~Elvira, F.M. de~Espinosa~Freijo, and
  J.~De~Vicente.
\newblock Using ultrasounds for the estimation of the misalignment in
  plate--plate torsional rheometry.
\newblock \emph{J. Phys. D: Appl. Phys.}, 46\penalty0 (20):\penalty0 205301,
  2013.

\bibitem[Savins and Metzner(1970)]{savins1970radial}
J.~G. Savins and A.~B. Metzner.
\newblock Radial (secondary) flows in rheogoniometric devices.
\newblock \emph{Rheol. Acta}, 9\penalty0 (3):\penalty0 365--373, 1970.

\bibitem[Singh et~al.(1999)Singh, Fogler, and Nagarajan]{singh1999prediction}
P.~Singh, H.S. Fogler, and N.~Nagarajan.
\newblock Prediction of the wax content of the incipient wax-oil gel in a
  pipeline: An application of the controlled-stress rheometer.
\newblock \emph{J. Rheol.}, 43\penalty0 (6):\penalty0 1437--1459, 1999.

\bibitem[Walters(1992)]{walters1992recent}
K.~Walters.
\newblock Recent developments in rheometry.
\newblock In \emph{Theoretical and applied rheology}, pages 16--23. Elsevier,
  1992.

\bibitem[Weller et~al.(1998)Weller, Tabor, Jasak, and
  Fureby]{weller1998tensorial}
H.~G. Weller, G.~Tabor, H.~Jasak, and C.~Fureby.
\newblock A tensorial approach to computational continuum mechanics using
  object-oriented techniques.
\newblock \emph{Comput. Phys.}, 12\penalty0 (6):\penalty0 620--631, 1998.

\end{thebibliography}
\end{document}